\documentclass[12pt]{article}
\usepackage{epsf,latexsym}
\usepackage{amsmath,amsfonts,amssymb,amsthm,cite}

\theoremstyle{definition}                    
 
\theoremstyle{remark}

\numberwithin{equation}{section}             

\usepackage[ps,dvips,arc,frame]{xy}

\newcommand{\rxyg}[2]{{\begin{xy} 0;<2mm,0mm>:<0mm,2mm>::0;0,
,(5,-2)*{a} ,(10,-1.8)*{b} ,(15,-2)*{c} ,(20,-1.8)*{d} ,(2,-5)*{a}
,(1.8,-10)*{b} ,(2,-15)*{c} ,(1.8,-20)*{d} ,(5,-5)*\cir(#1,0){}
,(10,-5)*\cir(#1,0){} ,(15,-5)*\cir(#1,0){} ,(20,-5)*\cir(#1,0){}
,(5,-10)*\cir(#1,0){} ,(10,-10)*\cir(#1,0){} ,(15,-10)*\cir(#1,0){}
,(20,-10)*\cir(#1,0){} ,(5,-15)*\cir(#1,0){} ,(10,-15)*\cir(#1,0){}
,(15,-15)*\cir(#1,0){} ,(20,-15)*\cir(#1,0){} ,(5,-20)*\cir(#1,0){}
,(10,-20)*\cir(#1,0){} ,(15,-20)*\cir(#1,0){} ,(20,-20)*\cir(#1,0){}
#2\end{xy}}}

 \newcommand{\rxyh}[2]{{\begin{xy} 0;<2mm,0mm>:<0mm,2mm>::0;0,
,(5,-2)*{a}
,(10,-2)*{\bar{a}} 
,(15,-1.8)*{b} 
,(20,-2)*{c} 
,(25,-1.8)*{d} 
,(30,-1.8)*{\bar{d}} 
,(2,-5)*{a}
,(2,-10)*{\bar{a}}
,(1.8,-15)*{b} 
,(2,-20)*{c} 
,(1.8,-25)*{d} 
,(1.8,-30)*{\bar{d}}
,(5,-5)*\cir(#1,0){}
,(10,-5)*\cir(#1,0){} 
,(15,-5)*\cir(#1,0){} 
,(20,-5)*\cir(#1,0){}
,(25,-5)*\cir(#1,0){} 
,(30,-5)*\cir(#1,0){}
,(5,-10)*\cir(#1,0){} 
,(10,-10)*\cir(#1,0){} 
,(15,-10)*\cir(#1,0){}
,(20,-10)*\cir(#1,0){} 
,(25,-10)*\cir(#1,0){}
,(30,-10)*\cir(#1,0){} 
,(5,-15)*\cir(#1,0){} 
,(10,-15)*\cir(#1,0){}
,(15,-15)*\cir(#1,0){} 
,(20,-15)*\cir(#1,0){} 
,(25,-15)*\cir(#1,0){}
,(30,-15)*\cir(#1,0){} 
,(5,-20)*\cir(#1,0){}
,(10,-20)*\cir(#1,0){} 
,(15,-20)*\cir(#1,0){} 
,(20,-20)*\cir(#1,0){}
,(25,-20)*\cir(#1,0){}
,(30,-20)*\cir(#1,0){} 
,(5,-25)*\cir(#1,0){}
,(10,-25)*\cir(#1,0){} 
,(15,-25)*\cir(#1,0){} 
,(20,-25)*\cir(#1,0){}
,(25,-25)*\cir(#1,0){}
,(30,-25)*\cir(#1,0){} 
,(5,-30)*\cir(#1,0){}
,(10,-30)*\cir(#1,0){} 
,(15,-30)*\cir(#1,0){} 
,(20,-30)*\cir(#1,0){}
,(25,-30)*\cir(#1,0){}
,(30,-30)*\cir(#1,0){} 
#2\end{xy}}}

\epsfverbosetrue
\newcommand{\T}{{\rm tr}}

\newcommand{\bb}{\begin{eqnarray}}
\newcommand{\ee}{\end{eqnarray}}
\newcommand{\eee}{\nonumber\end{eqnarray}}
\newcommand{\pp}[1]{\begin{pmatrix} #1 \end{pmatrix}}

\begin{document}

\thispagestyle{empty}

\begin{center}
CENTRE DE PHYSIQUE TH\'EORIQUE \footnote{\, Unit\'e Mixed de
Recherche (UMR) 6207 du CNRS et des Universit\'es Aix-Marseille 1 et 2 \\ \indent \quad \, Sud Toulon-Var, Laboratoire affili\'e \`a la 
FRUMAM (FR 2291)} \\ CNRS--Luminy, Case 907\\ 13288 Marseille Cedex 9\\
FRANCE
\end{center}

\vspace{1cm}

\begin{center}
{\Large\textbf{Almost-Commutative Geometry, massive Neutrinos
and the Orientability Axiom in $KO$-Dimension 6}} 
\end{center}

\vspace{1cm}

\begin{center}
{\large  Christoph A. Stephan$\,\,^{1,}$
\footnote{\, stephan@cpt.univ-mrs.fr} }

\vspace{1.5cm}

{\large\textbf{Abstract}}
\end{center}

In recent publications Alain Connes \cite{connes6} and John Barrett
\cite{barrett6} proposed to change the $KO$-dimension of the internal
space of the standard model in its noncommutative representation
\cite{cc} from zero to six. This apparently minor modification
allowed to resolve the fermion doubling problem \cite{testard},
and the introduction of Majorana mass terms for the right-handed 
neutrino. The price which had to be paid was that at least the
orientability axiom of noncommutative geometry  \cite{book,grav} may not be
obeyed by the underlying geometry.

In this publication we review three internal geometries, 
all three failing to meet the orientability axiom of noncommutative
geometry. They will serve as examples to illustrate the nature
of this lack of orientability. We will present an extension of 
the minimal standard model found in \cite{class6} by a right-handed
neutrino, where only the sub-representation associated to this
neutrino is not orientable. 

\vspace{1cm}

\vskip 1truecm

\noindent CPT-P80-2006\\
PACS-92: 11.15 Gauge field theories\\
\indent MSC-91: 81T13 Yang-Mills and other gauge theories

\vskip 1truecm

\noindent October 2006
\vskip 0.5truecm
\noindent \\
\noindent

\newpage

\section{Introduction}

During the last two decades Alain Connes developed noncommutative geometry
\cite{book},  which allows to unify two of the basic theories of modern physics: general relativity and the standard model of particle physics as classical field theories
\cite{cc}. In the noncommutative framework
the Higgs boson, which had previously to be put in by hand,  and many of the ad hoc features of the standard model appear in a  natural way. 

The basic geometric notion in the noncommutative framework is a spectral
triple, a set consisting of an algebra, a Hilbert space, a Dirac operator, a chirality
operator and a real structure, which corresponds the the charge conjugation
from particle physics \cite{book}. For a pedagogical  introduction see
\cite{schuck}.

For the unification of the  standard model and general relativity one finds that the
spectral triple is the tensor product of an internal geometry for the
standard model and a continuous geometry for space time. 
In this notion of geometry two different notions of dimension appear.
On the one hand the metric or spectral dimension which is given
by the behaviour of the eigenvalues of the Dirac operator. For space
time the relevant Dirac operator is just the ordinary Dirac operator
on curved space time, so one finds the metric dimension to be four.
The internal Dirac operator consists of the Fermionic mass matrix,
which has a finite number of eigenvalues, thus the internal metric
dimension is zero.

A second dimension which is assigned to each spectral triple is
the $KO$-dimension, an algebraic dimension with has its roots in 
K-theory. Until recently \cite{Dabrowski} it was suspected that the metric dimension
and the $KO$-dimension should be equal. Especially the 
internal space was taken to be of $KO$-dimension zero.
But John Barrett \cite{barrett6} and Alain Connes 
\cite{connes6} propose to depart from this paradigm, although
with different initial motivations, and to take the internal space of the standard
model to be of $KO$-dimension six.

This change of $KO$-dimension has several physical advantages.
It resolves the fermion doubling \cite{testard} problem by allowing to project out
the unphysical degrees of freedom resting in the internal space.
Furthermore Majorana masses could be introduced, if the 
orientability axiom \cite{book,grav} is weakened, in the sense
that it does not apply to the right-handed leptonic singlets. 

In this article we will present three internal or matrix geometries. They
all exhibit the general features of the  model with three summands
in the matrix algebra 
presented in \cite{barrett6,connes6}. All three have right-handed
Fermions which do explicitly violate the axiom of orientability.

The first example is a model of electro-weak type with two summands
in the matrix algebra. It has its origin 
in the early days of noncommutative geometry \cite{Dubois,cessay} and it also
appeared in the classification of the spectral
triples corresponding to internal spaces  in 
$KO$-dimension zero \cite{1,2,3,4} and $KO$-dimension six \cite{class6}.
It will serve as a toy model to illustrate the general features 
of the lack of orientability.

The second example will be the standard model in its version with three summands
summands in the matrix algebra \cite{real}.
Here the right-handed
leptonic sector is not orientable. Furthermore in $KO$-dimension
six \cite{connes6,barrett6} the model admits
a Majorana mass for the right-handed neutrinos and Lepto-Quark mass 
terms, connecting particles to antiparticles. The Lepto-Quark terms
are unwanted and have to be put to zero by hand \cite{barrett6}, or by imposing
extra conditions on the Dirac operator \cite{connes6}.

As a third example we will present an extension of the minimal standard
model with four summands in the matrix algebra which was found in \cite{4,class6}. 
This minimal model, containing
at least one massless neutrino, is extended by  
a right-handed neutrino with a Majorana mass term. It should be noted that
the minimal standard model obeys all the axioms of noncommutative
geometry.   It becomes only necessary to violate the axiom of orientability
when right-handed neutrinos are introduced in all three
generations.
The extension of the minimal standard model is based on its 
counterpart in $KO$-dimension zero presented in \cite{neutrmass}.
A further advantage of this model is that no Lepto-Quark terms are
allowed.

We will end this publication with a few speculations on the relationship
of the physical properties of the right-handed neutrino with Majorana 
mass term and the axiom of orientability.  
Furthermore we will give some remarks on the possibility of models
beyond the particle content of the standard model, especially the
$AC$-model presented in \cite{5}.

\section{Basic Definitions}

In this section we will give the necessary basic definitions  for finite noncommutative
geometries with $K0$-dimension six \cite{connes6,barrett6}. 
We restrict ourselves to real, finite spectral triples
($\mathcal{A},\mathcal{H},\mathcal{D}, $ $J,\chi$). The algebra $\mathcal{A}$ is
a finite sum of matrix algebras
$\mathcal{A}= \oplus_{i=1}^{N} M_{n_i}(\mathbb{K}_i)$ with $\mathbb{K}_i=\mathbb{R},\mathbb{C},\mathbb{H}$ where $\mathbb{H}$
denotes the quaternions. 
A faithful representation $\rho$ of $\mathcal{A}$ is given on the finite dimensional Hilbert space $\mathcal{H}$.
The Dirac operator $\mathcal{D}$ is a selfadjoint operator on $\mathcal{H}$ and plays the role of the fermionic mass matrix.
$J$ is an antiunitary involution, $J^2=1$, and is interpreted as the charge conjugation
operator of particle physics.
The chirality $\chi$  is a unitary involution, $\chi^2=1$, whose eigenstates with eigenvalue
$+1$ are interpreted as right-handed particle states and left-handed antiparticle
states, whereas  the eigenstates with eigenvalue $-1$  represent the left-handed
particle states and right-handed antiparticle states.
These operators are required to fulfill Connes' axioms for spectral triples:

\begin{itemize}
\item  
$[J,\mathcal{D}]=\{J,\chi \}=0,$ $ \mathcal{D}\chi =-\chi \mathcal{D}$, 

$[\chi,\rho(a)]=[\rho(a),J\rho(a')J^{-1}]= [[\mathcal{D},\rho(a)],J\rho(a')J^{-1}]=0, \forall a,a' \in \mathcal{A}$.

Note the change of the commutator $[J,\chi]=0$ from $KO$-dimension zero 
to the anti-commutator $\{J,\chi \}=0$ in $KO$-dimension six.
\item The intersection form
$\cap_{ij}:=\T(\chi \,\rho (p_i) J \rho (p_j) J^{-1})$ is non-degenerate,
$\rm{det}\,\cap\not=0$. The
$p_i$ are minimal rank projections in $\mathcal{A}$. This condition is called
{\it Poincar\'e duality}. Demanding the Poincar\'e duality to hold requires
an even number of summands in the matrix algebra \cite{barrett6,class6}. 
\item The chirality can be written as a finite sum $\chi =\sum_i\rho(a_i)J\rho(a'_i)J^{-1}$,
which is a $0$-dim Hochschild cycle.
This condition is called {\it orientability}. 
\end{itemize} 
As mentioned in the introduction it will be necessary to  weaken the last axiom
since right-handed neutrinos with Majorana masses cannot fulfil the orientability
axiom.  

\section{The Models}

In this section we will treat three matrix geometries which will not in general obey
all the axioms of finite spectral triples stated above. The first model which we
present is an electro-weak model based on the geometry with two summands
in the matrix algebra \cite{Dubois,cessay}.  It is the purpose of
this model to clarify why the orientability axiom does not hold if a Majorana mass
term is desired.

The second model is the
standard model with three summands in the matrix algebra \cite{real} in
its version with $KO$-dimension six \cite{connes6,barrett6}. 
For this model the Poincar\'e duality has to be slightly
modified. Instead of being valid for the spectral triple in its whole, the leptonic
sector and the quark sector are treated separately and for each sector
the Poincar\'e duality holds  \cite{connespriv}. 

The last model
will be based on the standard model in its formulation with four summands
in the matrix algebra. This model appeared for the first time in the four summand
case with $KO$-dimension zero \cite{4} and again in the classification for
the case of $KO$-dimension six \cite{class6}. We will enlarge the particle content
 following \cite{neutrmass}, by introducing a right-handed neutrino and additionally
furnishing it with a Majorana mass.

\subsection{The Electro-Weak Model}

We will start with the electro-weak model with two summands in the 
matrix algebra, \cite{Dubois,cessay}, which also appeared in the classification
\cite{1}. The geometry is furnished by a matrix algebra with two summands
$\mathcal{A}_{EW} = \mathbb{C} \oplus M_2 (\mathbb{C})\ni (a,b)$. It has been 
shown that the physical models based on this geometry suffer from
Yang-Mills-anomalies \cite{4}, but nonetheless it will serve us as a toy model
to exemplify some basic properties connected the orientability axiom in
$KO$-dimension six.

We take the the left-(right-) handed sub-representation to be
\bb
\rho _L (a,b) = b,
\quad
\rho _R (a,b) = \bar{a},
\quad
\rho_L^c (a,b) = a 1_2,
\quad
\rho _R^c (a,b,c) =a,
\ee
where $1_2$ denotes the $2$-dim. unity matrix.
The complete representation is given by 
\bb
\rho (a,b)  = \rho _L (a,b) \oplus \rho _R (a,b) \oplus  \overline{\rho _L^c} 
(a,b) \oplus \overline{\rho _R^c} (a,b)
\label{rep}
\ee
The  left- and right-handed particle and antiparticle subspaces
\bb
\mathcal{H}^{PL} =  \pp{ e \\ \nu}^L, 
\quad 
\mathcal{H}^{PR} = \nu^R, 
\quad 
\mathcal{H}^{AL} =  \pp{  e^c \\ \nu^c}^L, \quad
\mathcal{H}^{AR} =  \nu^{cR}.
\ee
build the complete Hilbert space  given by
$\mathcal{H} = \mathcal{H}^{PL} \oplus \mathcal{H}^{PR} \oplus \mathcal{H}^{AL} \oplus \mathcal{H}^{AR}$. Note that this notation can lead to confusion because $\nu^{cR}$
is the charge conjugate of $\nu^{R}$. Since chirality and charge conjugation
anti-commute $\nu^{cR}$ is actually left-handed, confirm \ref{chi2}.
It follows from the axioms \cite{connes6,barrett6,class6} that the Dirac operator
has the general form
\bb
\mathcal{D} = \pp{\Delta & H \\ H^* & \bar{\Delta} },
\label{Dirac}
\ee
with $\Delta=\Delta^*$ being a complex matrix, connecting left-handed to right-handed particles
and vice versa, and $H$ being a complex symmetric matrix connecting
particles to antiparticles. In the basis defined
by the Hilbert space, $\Delta$ can thus be
written
\bb
\Delta = \pp{0 & M \\ M^* & 0},
\label{Dirac}
\ee
with $M$ a complex $1 \times 2$ matrix.
To be consistent with the first order axiom $ [[\mathcal{D},\rho(a)],J\rho(a')J^{-1}]=0,$    
for all $a,a' \in \mathcal{A}_{EW}$ we find that the matrix connecting particles
and antiparticles has to take the form
\bb
H = \pp{0 & 0\\ 0 & M_{\nu \bar{\nu}}}.
\ee
$ M_{\nu \bar{\nu}}$ is a complex number, the Majorana mass of the
neutrino. Note that a corresponding Majorana mass for a right handed
electron would not be possible. The electro-weak model with a right-handed electron
at the place of the right-handed neutrino has as its algebra representation
\bb
\tilde{\rho}_L (a,b) = b,
\quad
\tilde{\rho}_R (a,b) = a,
\quad
\tilde{\rho}_L^c (a,b) = \bar{a} 1_2,
\quad
\tilde{\rho}_R^c (a,b,c) =\bar{a}.
\label{elec}
\ee
Here the first order axiom demands $H$ to be identically zero.

Let us return to the electro-weak model with one right-handed neutrino.
We can  write down the real structure
\bb
J = \pp{0 & 1 \\ 1 & 0} \; \circ \; complex \; conjugation
\label{real}
\ee
which is nothing but the charge conjugation operator from particle physics.
With the chirality 
\bb
\chi = {\rm diag}(-1_2,1,1_2,-1)
\label{chi2}
\ee
and the fact that the algebra $\mathcal{A}_{EW}$ has only two 
projectors $p_1=(1,0)$ and $p_2=(0,1_2)$ it is easy to check
that the Poincar\'e duality holds. The same is true for all other
axioms safe the orientability.

For the orientability we need to be able to write $\chi$ as a $0$-dim.
Hochschild cycle, i.e.
\bb
\chi = \sum_i\rho(a_i)J\rho(a'_i)J^{-1}.
\ee
But, taking the representation (\ref{rep}) and the real structure (\ref{real})
we find
\bb
\sum_i \pp{ b_i & 0 &0 &0 \\ 0 & a_i&0&0 \\ 0&0&\bar{a}_i 1_2 &0 \\ 0&0&0& \bar{a}_i}  \pp{ a'_i 1_2 & 0 &0 &0 \\ 0 & a'_i&0&0 \\ 0&0& b'_i &0 \\ 0&0&0& \bar{a}'_i}
&=& \sum_i \pp{  b_i a'_i & 0 &0 &0 \\ 0 & a_i a'_i &0&0 \\ 0&0&\bar{a}_i b'_i &0 \\ 0&0&0& \bar{a}_i \bar{a}'_i} \nonumber \\ \\
&\neq& \pp{ -1_2& 0 &0 &0 \\ 0 & 1&0&0 \\ 0&0& 1_2 &0 \\ 0&0&0& -1} = \chi
\nonumber 
\ee
since the matrix algebra is taken to be an algebra over the real numbers. 
It follows that the
orientability axiom does not hold. This is a general feature of representations
where the particle sub-representation for a given particle multiplet consists
of elements stemming from the same sub-matrix algebra as the sub-representation
from the corresponding antiparticle multiplet. 
Here the case for the
right-handed neutrino and its antiparticle is treated but the same is true if one takes
the model with a right-handed electron (\ref{elec}). For a general treatment
we refer to \cite{class6}

\subsection{The Three Summands Model}

We will present the the standard model with three summands in the matrix 
algebra \cite{real} and one generation of Fermions. 
As in the electro-weak case the $KO$-dimension will be taken to 
be six \cite{connes6,barrett6}.
The second and
third generation of Fermions can be introduced by simply copying the algebraic
set-up of the first generation. This also allows to introduce CKM-mixing matrices
for the quarks and the leptons \cite{schuck,connes6}.

The internal algebra is chosen to be  $\mathcal{A}_{3} = \mathbb{C} \oplus \mathbb{H} \oplus M_3(\mathbb{C})  \ni (a,b,c)$ and the representation of the algebra, ordered into left and right, particle and antiparticle part is given by:
\bb
\rho _L (a,b,c) &=&\pp{b\otimes 1_3&0\\ 0& b},\quad
\rho _R (a,b,c) =\pp{a 1_3&0&0&0\\ 0& \bar{a} 1_3&0&0\\ 0&0& a&0 \\ 0&0&0&\bar{a}},
\nonumber \\
\nonumber \\
\rho _L^c (a,b,c) &=&\pp{1_2\otimes c&0\\ 0& \bar{a} 1_2},\quad
\rho _R^c (a,b,c) =\pp{c&0&0&0\\ 0&c&0&0\\ 0&0& \bar{a}& 0 \\ 0&0&0&\bar{a}},
\ee
where $1_3$ is the unit matrix and the complex conjugates were chosen
in order to reproduce the standard model. The complete representation is defined
as in (\ref{rep}).
The Hilbert space is copied from particle physics with the usual Fermion multiplets. 
One has for the left- and right-handed particle subspaces
\bb
\mathcal{H}^{PL} = \pp{\pp{d \\ u}^L \\ \pp{ \nu_e\\ e}^L }, \quad 
\mathcal{H}^{PR} = \pp{ d^R \\ u^R  \\ \nu_e^R \\ e^R} 
\label{H1}
\ee
and for the left- and right-handed antiparticle subspaces
\bb
\mathcal{H}^{AL} = \pp{\pp{d^c \\ u^c}^L \\ \pp{  \nu_e^c \\ e^c}^L}, \quad
\mathcal{H}^{AR} = \pp{ d^{cR} \\ u^{cR}  \\ \nu_e^{cR}\\ e^{cR}}.
\label{H2}
\ee
The complete Hilbert space is then given by
$\mathcal{H} = \mathcal{H}^{PL} \oplus \mathcal{H}^{PR} \oplus \mathcal{H}^{AL} \oplus \mathcal{H}^{AR}$.

The particle part $\Delta$ of the Dirac operator with the general form (\ref{Dirac}) is
\bb
\Delta = \pp{0&0& M_d \otimes 1_3 & M_u \otimes 1_3 &0&0 \\
0 &0&0&0& M_{\nu_e} & M_{e}\\
M_d^{\ast} \otimes 1_3 &0&0&0&0 &0\\
M_u^{\ast} \otimes 1_3 &0&0&0&0&0 \\
0&M_{\nu_e}^{\ast} &0&0&0&0 \\
0&M_{e}^{\ast} &0&0&0&0}
\label{Delta}
\ee
where the  four mass matrices $M_d$, $M_u$, $M_{\nu_e}$
and $M_{e}$ are $2\times 1$ complex matrices which may 
be chosen conveniently.

For the Majorana part $H$ of the Dirac operator the most general form
is
\bb
H = \pp{0&0& 0 & 0 &0&0 \\
0 &0&0&0& 0& 0\\
0&0&0&0&0 &0\\
0&0&0&0&M_{u \bar{\nu}}&M_{u\bar{e}} \\
0&0&0&M_{u \bar{\nu}}^T&M_{\nu \bar{\nu}}&0 \\
0&0&0&M_{u\bar{e}}^T&0&0},
\ee
where the Lepto-Quark matrices $M_{u\bar{e}}$ and $M_{u \bar{\nu}}$ are
complex $3 \times 1$ matrices and the Majorana mass term $M_{\nu \bar{\nu}}$
of the right-handed neutrino is a complex number. The matrix $H$ is symmetric
as desired. It is possible to eliminate the unwanted mass terms mixing leptons
and quarks be demanding the Dirac operator to commute with $(a,a 1_2,0) \in 
\mathcal{A}_{3}$, \cite{connes6}.

From the electro-weak example we learn, that the geometry based on 
the algebra $\mathcal{A}_{3}$  does not obey the  axiom of 
orientability in the right-handed leptonic sector \cite{connespriv}. I seems especially
worrisome that the right-handed electron does not furnish a sub-representation
which obeys the requirements of the orientability. 

We will  propose
a model in which the particles of the minimal standard model meet the
requirements of  the orientability axiom \cite{class6} and only the right-handed neutrino
does not. By minimal standard model we mean the standard model with 
only a left-handed neutrino in the first generation. We will see that this model 
also allows for a Majorana mass term.

\subsection{The Four Summands Model}

As a basis for the first generation of the standard model with a right-handed neutrino 
and its Majorana mass we will take the minimal standard model based on the
Krajewski diagram depicted in figure 1
\begin{center}
\begin{tabular}{c}
\rxyg{0.7}{
,(10,-15)*\cir(0.3,0){}*\frm{*}
,(10,-15);(5,-15)**\dir2{-}?(.6)*\dir{>}
,(10,-20);(5,-20)**\dir{-}?(.6)*\dir{>}
} \\ \\
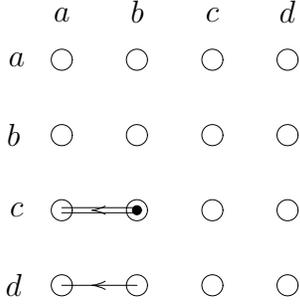
Fig. 1: Krajewski diagram of the minimal standard model.
\\
\end{tabular}
\end{center}
which is the diagram 12 
in \cite{class6} with the first two summands exchanged. 
Here the quark sector is encoded in the double arrow and the
lepton sector in the single arrow. For a detailed guide for translation of 
Krajewski diagrams to spectral triples we refer to \cite{Kraj,1} and \cite{class6}.
The minimal standard model 
with four summands in the matrix algebra fulfils all the axioms of non-commutative
geometry, including the orientability axiom. 
One should note the similarity of this model to the Connes-Lott model \cite{cl}.

Following  \cite{neutrmass}, the Hilbert space
and consequently the representation and the Dirac operator will be enlarged. 
We give the corresponding Krajewski diagram in figure 2, from which the representation
and the Dirac operator can be read off. Note that in figure 2 the particles and the
antiparticles are depicted. 
\begin{center}
\begin{tabular}{c}
\rxyh{0.7}{
,(15,-20)*\cir(0.3,0){}*\frm{*}
,(15,-25)*\cir(0.3,0){}*\frm{*}
,(15,-20);(10,-20)**\dir{-}?(.6)*\dir{>}
,(15,-20);(5,-20)**\crv{(10,-17)}?(.6)*\dir{>}
,(15,-25);(10,-25)**\dir{-}?(.6)*\dir{>}
,(15,-25);(30,-25)**\crv{(22.5,-22)}?(.55)*\dir{>}
,(20,-15)*\cir(0.3,0){}*\frm{*}
,(25,-15)*\cir(0.3,0){}*\frm{*}
,(20,-10);(20,-15)**\dir{-}?(.6)*\dir{>}
,(20,-5);(20,-15)**\crv{(17,-10)}?(.6)*\dir{>}
,(25,-10);(25,-15)**\dir{-}?(.6)*\dir{>}
,(25,-30);(25,-15)**\crv{(22.5,-22.5)}?(.6)*\dir{>}
,(30,-25)*\cir(0.3,0){}*\frm{*}
,(25,-30)*\cir(0.3,0){}*\frm{*}
,(25,-30);(30,-25)**\dir{--}?(.6)*\dir{>}
} \\ \\
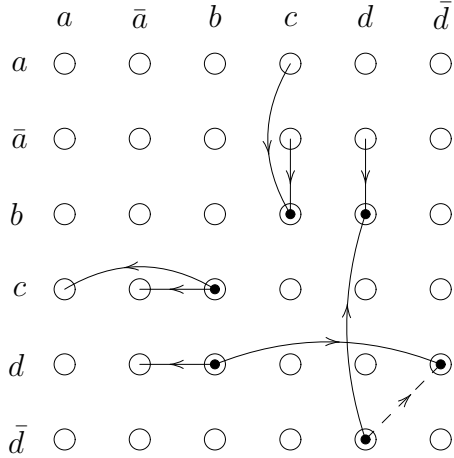
Fig. 2: Krajewski diagram of the standard model with right-handed neutrino \\
\hskip-1.1cm and Majorana-mass term depicted by the dashed arrow. 
\\
\end{tabular}
\end{center}
The particle mass terms are represented by the horizontal arrows, the antiparticle
mass terms by the vertical arrows and the Majorana mass term, which connects
the right-handed neutrino with its charge conjugate, by a dashed diagonal arrow.  
One should perhaps also note the aesthetic appeal of this Krajewski diagram.

The
only axioms that will be violated in this model is the orientability axiom, 
i.e. all the other axioms of noncommutative geometry
still hold. Furthermore the Majorana mass term is now permitted by the first order
axiom. Note, that Lepto-Quark-terms are prohibited by the same axiom.
 
The internal algebra is chosen to be  $\mathcal{A}_{4} = \mathbb{C} \oplus \mathbb{H} \oplus M_3(\mathbb{C}) \oplus \mathbb{C} \ni (a,b,c,d)$ and the representation of the algebra, ordered into left and right, particle and antiparticle part is given by:
\bb
\rho _L (a,b,c,d) &=&\pp{b\otimes 1_3&0\\ 0& b},\quad
\rho _R (a,b,c,d) =\pp{a 1_3&0&0&0\\ 0& \bar{a} 1_3&0&0\\ 0&0& \bar{d} &0 \\ 0&0&0&\bar{a}},
\nonumber \\
\nonumber \\
\rho _L^c (a,b,c,d) &=&\pp{1_2\otimes c&0\\ 0& d 1_2},\quad
\rho _R^c (a,b,c,d) =\pp{c&0&0&0\\ 0&c&0&0\\ 0&0& d& 0 \\ 0&0&0&d},
\label{rep4}
\ee
where the complex conjugates where chosen
in order to reproduce the standard model. The complete representation is then
defined as in the case of the three summand model, as the direct sum of the
particle representations and the complex conjugates of the antiparticle representations.
It acts on the Hilbert
space $\mathcal{H} = \mathcal{H}^{PL} \oplus \mathcal{H}^{PR} \oplus \mathcal{H}^{AL} \oplus \mathcal{H}^{AR}$
where the subspaces are copied from the three summand model (\ref{H1}), (\ref{H2}).
Note furthermore that the particle and antiparticle sub-representations of the 
right-handed electron contain elements from different summands of the matrix
algebra $\mathcal{A}_{4}$:
\bb
\rho_{e,R} = \bar{a}, \quad \rho_{e,R}^c = d.
\ee
Therefore this sub-representation obeys the orientability axiom.
This should be contrasted with the  sub-representations of the right-handed
neutrino,
\bb
\rho_{\nu,R} = \bar{d}, \quad \rho_{\nu,R}^c = d,
\ee
which violates the orientability axiom.

The particle part $\Delta$ of the Dirac operator coincides with the three summand
model (\ref{Delta}). But the sub-matrix $H$ differs considerably since the representation
(\ref{rep4}) prohibits mass terms mixing leptons and quarks. It allows 
a Majorana mass term for the right-handed neutrino,
\bb
H = \pp{0&0& 0 & 0 &0&0 \\
0 &0&0&0& 0& 0\\
0&0&0&0&0 &0\\
0&0&0&0&0&0 \\
0&0&0&0&M_{\nu \bar{\nu}}&0 \\
0&0&0&0&0&0}.
\ee
It is of course straight forward to enlarge this model to three generations of fermions. 
This allows to introduce a CKM-matrix for the quark mixing a and a leptonic mixing
matrix.  Note that this Majorana mass  term of the right-handed neutrino
is gauge invariant and therefore obeys the first order axiom. Furthermore
it does not participate in the fluctuation of the Dirac operator and enters
in a different way into the spectral action than the other Dirac mass terms 
\cite{connes6}.

One should furthermore point out, that minimal model allows to introduce massive,
right-handed neutrinos in the second and third Fermion generation, obeying
{\it all} axioms of non-commutative geometry. But in the minimal case at least
one neutrino has to remain purely left-handed and massless. This model
is not in conflict with current experiments. It would be ruled out if the masses
of the neutrinos could be measured directly, and would all be found to be non-zero,
or if the neutrinoless double beta-decay would be observed. 

\section{Speculations}

How has one to interpret this violation of the orientability axiom which necessarily
appears, at least for right-handed neutrinos with Majorana masses? Since
the right-handed neutrino is completely sterile with respect to the gauge-group,
it does not possess electro-weak charges or colour charge, the chirality is
the only way to discriminate between particle and antiparticle. In the case
of $KO$-dimension zero this r\^ole was played by the $S^0$-real structure.
But from the physical point of view the right-handed neutrino may be considered
as its proper antiparticle if a Majorana mass term exists.  So there
is no difference between the particle and its antiparticle and one could
expect that the orientability axiom does not apply to this scenario.
The corresponding
physical process to decide over the Majorana nature of the 
right-handed electron-neutrino would be the neutrinoless double beta-decay.

Furthermore the Majorana mass $M_{\nu \bar{\nu}}$ is usually taken to
be of the order of the Planck mass. Therefore the right-handed neutrino,
$\nu_e^R$ would be extremely heavy, and therefore would not be
present in the low-energy. We do expect the
noncommutative geometry of the standard model to be a low-energy 
approximation which should be replaced by a different geometry at high 
energies close to the Planck mass. This high-energy geometry may
be able to include the right-handed neutrino into the complete noncommutative
geometric frame-work.

As a remark we would like to point out that the extension of the
standard model by new Fermions, baptised $AC$-Fermions \cite{5},
is compatible with the approach of $KO$-dimension six. These
particles originate from the combination of the standard model
and the electro-strong model presented in \cite{4,class6}. It allows
to enlarge the particle content by two oppositely electro-magnetically 
charged Fermions which may be suitable candidates for dark matter
\cite{Fargion:2005ep,Khlopov:2006uv}. These particles are as
well consistent with current high-energy experiments \cite{Knecht:2006tv}.
It is remarkable that this extension is among the few possible
physically sensible almost-commutative geometries found in \cite{class6}.

\section{Conclusion} 

Inspired by the recent articles of
Alain Connes \cite{connes6} and  John Barrett \cite{barrett6}  
we presented an illustration of the problems
this type of models has concerning the  orientability axiom of noncommutative
geometry. It turned out that especially right-handed neutrinos with
a Majorana mass term, a vital ingredient for the See-Saw-mechanism, 
do not obey this axiom.

We furthermore extended the minimal standard model with four summands
in the matrix algebra by a right-handed neutrino. This model suffers
of course from the same short-commings concerning the  orientability.
But we think one can argue that this effect may be considered as
a low energy effect, since the problematic parts of the representation
correspond to extremely heavy right-handed neutrinos, with a Majorana
mass of the order of the Planck mass. Furthermore the lack of orientability
of the right-handed neutrino may be linked to the physical point of view that
this particle is its own antiparticle. 

We are now in the very interesting situation that  experimental physics 
may give us deep insights into the internal geometry of space time. If the
neutrinoless double beta-decay would be experimentally confirmed, i.e.
the electron-neutrino would possess a Majorana mass, one would have
to consider one of the above spectral geometries as internal space.
This would exclude the case of $KO$-dimension zero, where Majorana
masses may not exist and would require to take geometries into account,
which do not respect the orientability axiom.

\vskip1cm
\noindent
{\bf Acknowledgements:} The author would like to thank T. Sch\"ucker for careful proof
reading. We
gratefully acknowledge a fellowship of the Alexander von Humboldt-Stiftung.

\bibliographystyle{unsrt}
\bibliography{noncom,kosmo}

\end{document}